\documentclass[letter,longauth,traditabstract]{aa} 
\usepackage{graphicx}
\usepackage{txfonts}
\usepackage{natbib}
\bibpunct{(}{)}{;}{a}{}{,}

\newcommand{\um}{\hbox{${\mu}$m}}
\newcommand{\Msun}{$M_{\sun}$}

\begin{document}

\title 
{
Filamentary structures and compact objects in the Aquila and Polaris clouds observed by \emph{Herschel}
\thanks{\emph{Herschel} is an ESA space observatory with science instruments provided by European-led Principal Investigator
consortia and with important participation from NASA.}
\thanks{Fig.~2 and Appendix A are only available in electronic form: \,\,\,\,\,\,\,\,\,\,\,\,\, http://www.edpsciences.org}    
}




\author
{
A.~Men'shchikov\inst{1} 
\and Ph.~Andr\'e\inst{1} 
\and P.~Didelon\inst{1}
\and V.~K\"onyves\inst{1} 
\and N.~Schneider\inst{1}
\and F.~Motte\inst{1}
\and\\S.~Bontemps\inst{1}
\and D.~Arzoumanian\inst{1}
\and M.~Attard\inst{1}\\
A.~Abergel\inst{2}
\and J.-P.~Baluteau\inst{3}
\and J.-Ph.~Bernard\inst{5}
\and L.~Cambr\'esy\inst{6}
\and P.~Cox\inst{7}
\and J.~Di~Francesco\inst{8}
\and A.~M.~di~Giorgio\inst{9}
\and M.~Griffin\inst{4}
\and P.~Hargrave\inst{4}
\and M.~Huang\inst{10}
\and J.~Kirk\inst{4}
\and J.~Z.~Li\inst{10}
\and P.~Martin\inst{11}
\and V.~Minier\inst{1}
\and M.-A.~Miville-Desch{\^e}nes\inst{2,11}
\and S.~Molinari\inst{9}
\and G.~Olofsson\inst{12}
\and S.~Pezzuto\inst{9}
\and H.~Roussel\inst{13}
\and D.~Russeil\inst{4}
\and P.~Saraceno\inst{9}
\and M.~Sauvage\inst{1}
\and B.~Sibthorpe\inst{14}
\and L.~Spinoglio\inst{9}
\and L.~Testi\inst{15}
\and D.~Ward-Thompson\inst{4}
\and G.~White\inst{16,17}
\and C.~D.~Wilson\inst{18}
\and A.~Woodcraft\inst{19}
\and A.~Zavagno\inst{4}
}


\institute
{
Laboratoire AIM, CEA/DSM--CNRS--Universit{\'e} Paris Diderot, IRFU/Service d'Astrophysique, C.E. Saclay, Orme des Merisiers, 91191 Gif-sur-Yvette, France 
\and Institut d’Astrophysique Spatiale (CNRS), Universit{\'e} Paris-Sud, b{\^a}t. 121, 91405, Orsay, France            
\and Laboratoire d'Astrophysique de Marseille, CNRS/INSU--Universit\'e de Provence, 13388 Marseille Cedex 13, France   
\and School of Physics and Astronomy, Cardiff University, Queens Buildings, The Parade, Cardiff CF24 3AA, UK           
\and CESR, 9 Avenue du Colonel Roche, B.P. 4346, F-31029 Toulouse, France                                              
\and CDS, Observatoire de Strasbourg, 11, rue de l'Universit{\'e}, 67000 Strasbourg, France                            
\and IRAM, 300 rue de la Piscine, Domaine Universitaire, 38406 Saint Martin d'H{\'e}res, France                        
\and Herzberg Institute of Astrophysics, University of Victoria, Department of Physics and Astronomy, Victoria, Canada 
\and INAF-IFSI, Fosso del Cavaliere 100, 00133 Roma, Italy                                                             
\and National Astronomical Observatories, Chinese Academy of Sciences, A20 Datun Road, Chaoyang District, Beijing 100012, China 
\and Canadian Institute for Theoretical Astrophysics, University of Toronto, 60 St. George Street, Toronto, ON, M5S 3H8, Canada 
\and Department of Astronomy, Stockholm University, AlbaNova University Center, SE-10691 Stockholm  
\and Institut d'Astrophysique de Paris, UMR7095 CNRS, Universit\'e Pierre \& Marie Curie, 98 bis Boulevard Arago, F-75014 Paris, France 
\and UK Astronomy Technology Centre, Royal Observatory Edinburgh, Blackford Hill, EH9 3HJ, UK       
\and Istituto Nationale di Astrofisica, Largo Enrico Fermi 5, I-50125 Firenze, Italy                
\and The Rutherford Appleton Laboratory, Chilton, Didcot OX11 0NL, UK                               
\and Department of Physics \& Astronomy, The Open University, Milton Keynes MK7 6AA, UK             
\and Dept. of Physics \& Astronomy, McMaster University, Hamilton, Ontario, L8S 4M1, Canada         
\and UK Astronomy Technology Center, Royal Observatory Edinburgh, Edinburgh, EH9 3HJ, UK            
}

\date{Received 1 April 2010; accepted 3 May 2010}

\offprints{Alexander Men'shchikov}
\mail{alexander.menshchikov@cea.fr}
\titlerunning{Filaments and objects in the Aquila and Polaris clouds}
\authorrunning{Men'shchikov et al.}


\abstract{

Our PACS and SPIRE images of the Aquila Rift and part of the Polaris Flare regions, taken during the science demonstration phase of
\emph{Herschel} discovered fascinating, omnipresent filamentary structures that appear to be physically related to compact cores.
We briefly describe a new multi-scale, multi-wavelength source extraction method used to detect objects and measure their
parameters in our \emph{Herschel} images. All of the extracted starless cores (541 in Aquila and 302 in Polaris) appear to form in
the long and very narrow filaments. With its combination of the far-IR resolution and sensitivity, \emph{Herschel} \emph{directly}
reveals the filaments in which the dense cores are embedded; the filaments are resolved and have deconvolved widths of $\sim$
35{\arcsec} in Aquila and $\sim$ 59{\arcsec} in Polaris ($\sim$ 9000\,AU in both regions). Our first results of observations with
\emph{Herschel} enable us to suggest that in general dense cores may originate in a process of fragmentation of complex networks of
long, thin filaments, likely formed as a result of an interplay between gravity, interstellar turbulence, and magnetic fields. To
unravel the roles of the processes, one has to obtain additional kinematic and polarization information; these follow-up
observations are planned.

}


\maketitle


\begin{figure*}
\centering
\centerline{\resizebox{0.5\hsize}{!}{\includegraphics{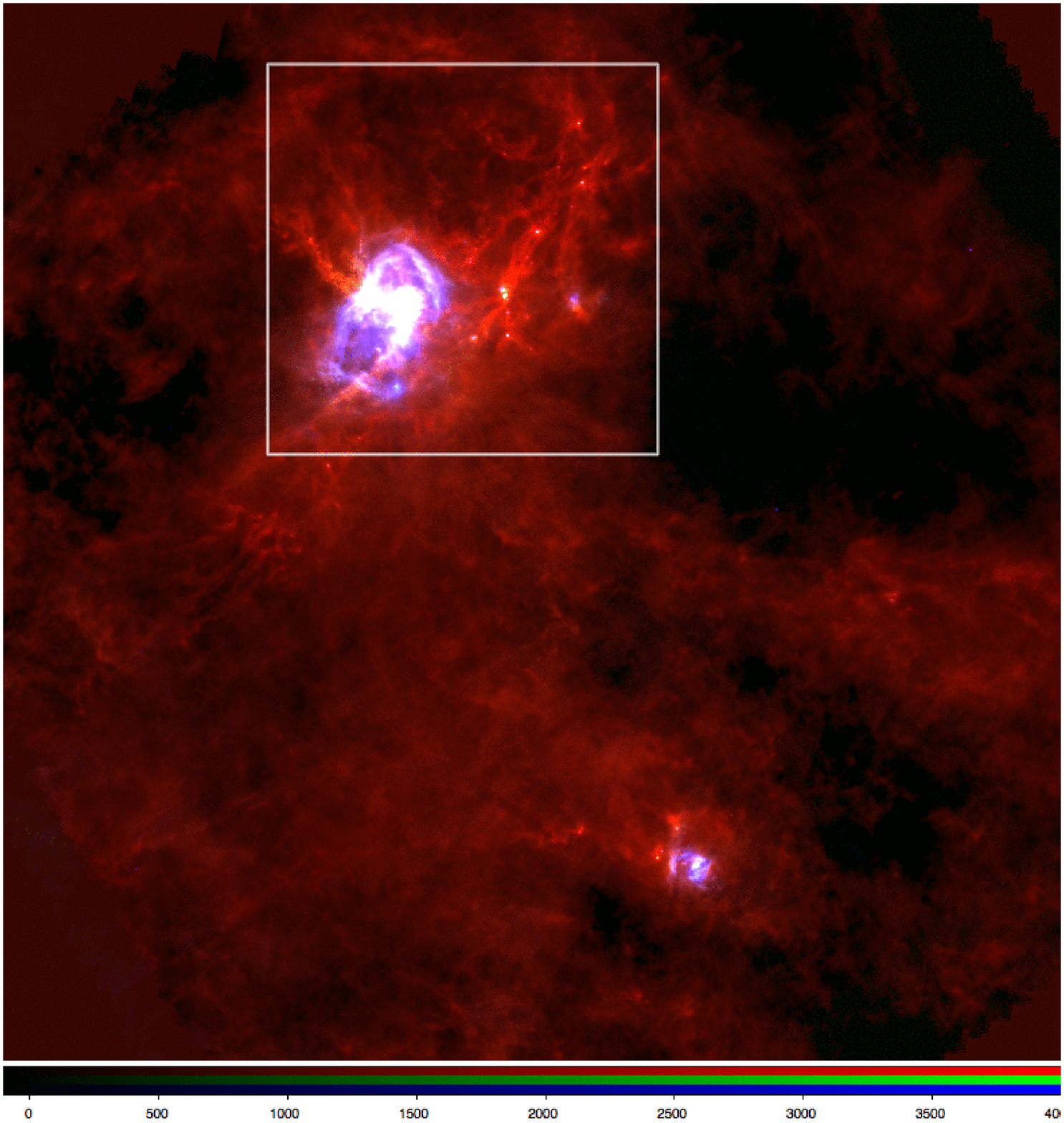}}
            \resizebox{0.5\hsize}{!}{\includegraphics{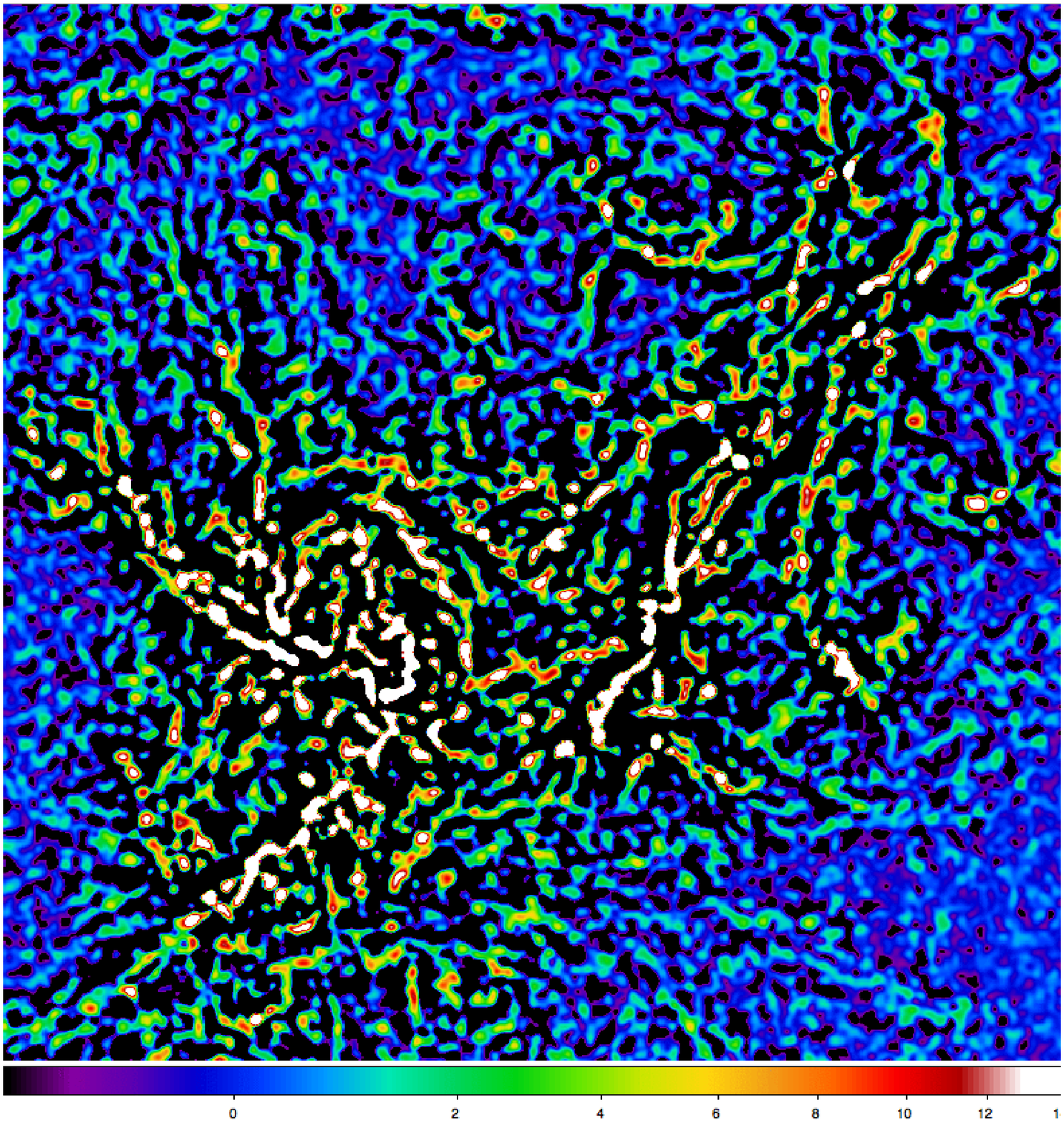}}}
\centerline{\resizebox{0.5\hsize}{!}{\includegraphics{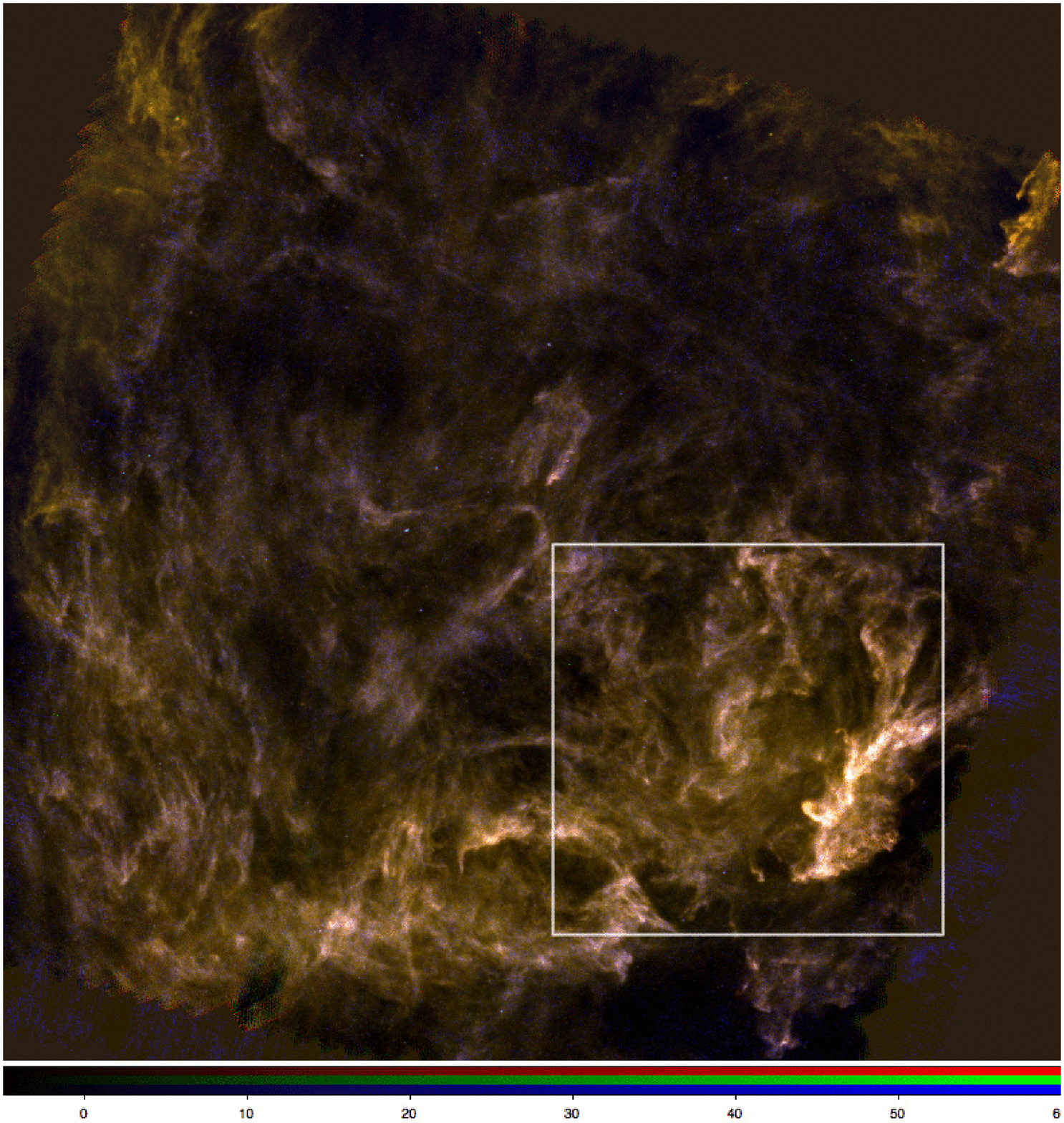}}
            \resizebox{0.5\hsize}{!}{\includegraphics{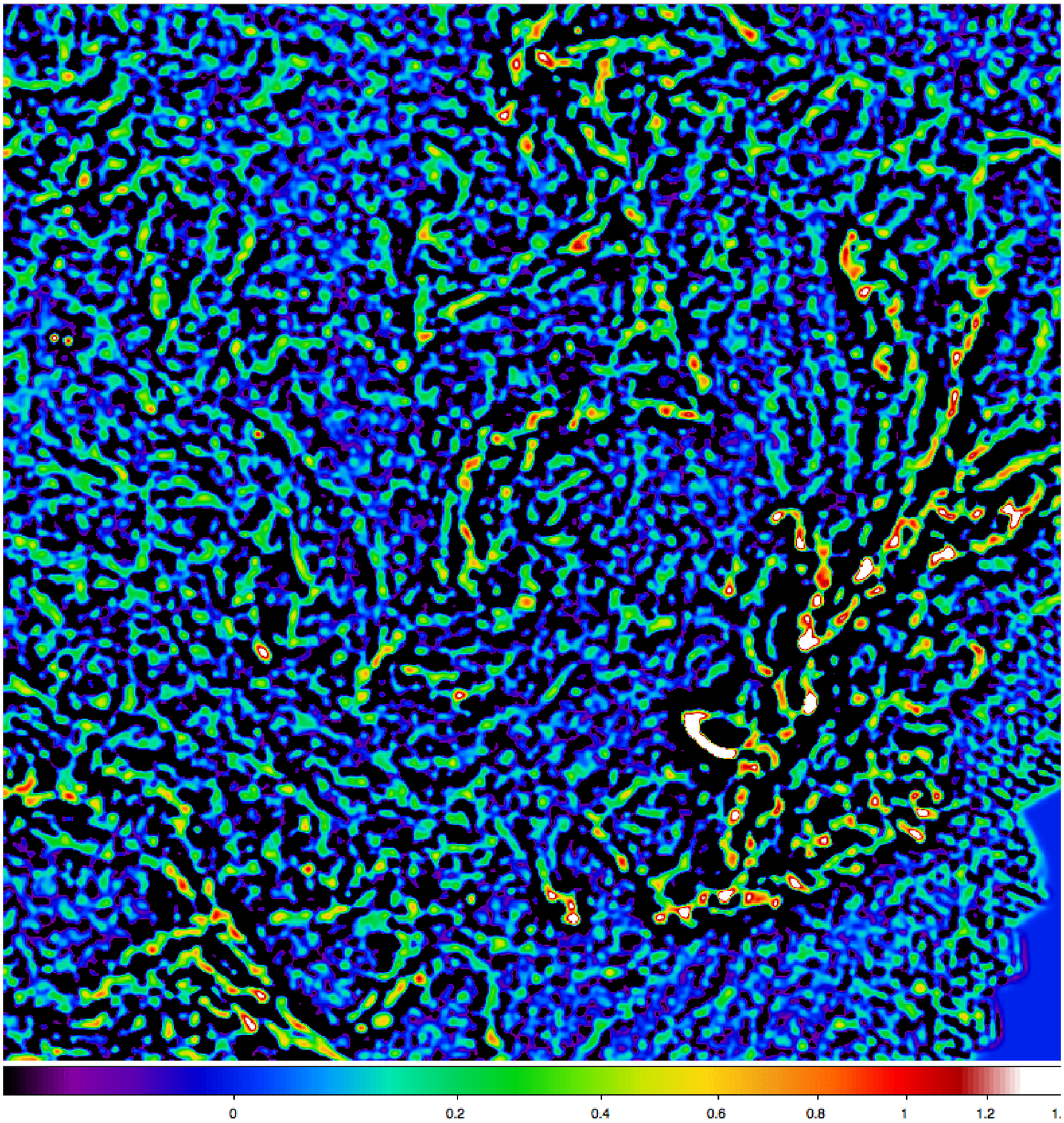}}}
\caption{Composite 3-color images (\emph{left}) produced from the observed images of the Aquila (\emph{top}) and Polaris 
         (\emph{bottom}) fields (panels are 3{\fdg}3$\times$3{\fdg}3, or $\sim$15 pc and $\sim$9 pc at their distances of 
         260 pc and 150 pc, respectively; see Sect.~\ref{results.and.discussion}); the red color (SPIRE band at 350\,{\um}), 
         highlights colder areas and objects, whereas the green and blue colors (PACS bands at 160\,{\um} and 70\,{\um}) show 
         regions with progressively hotter radiation fields. High-contrast ``single-scale'' decompositions (\emph{right}) of 
         sub-fields in Aquila (\emph{top}) and Polaris (\emph{bottom}) combined from those in all SPIRE bands for better visibility, 
         display intensity distribution (MJy/sr) within a narrow range of spatial scales around 40{\arcsec}; the sub-fields of 
         $1{\fdg}2\times 1{\fdg}2$ are outlined by white squares in the left panels. Clearly visible is a very tight association 
         of the narrow filaments with compact cores.}
\label{figure1}
\end{figure*}

\onlfig{2}
{
\begin{figure*} 
\centering
\centerline{\resizebox{0.5\hsize}{!}{\includegraphics{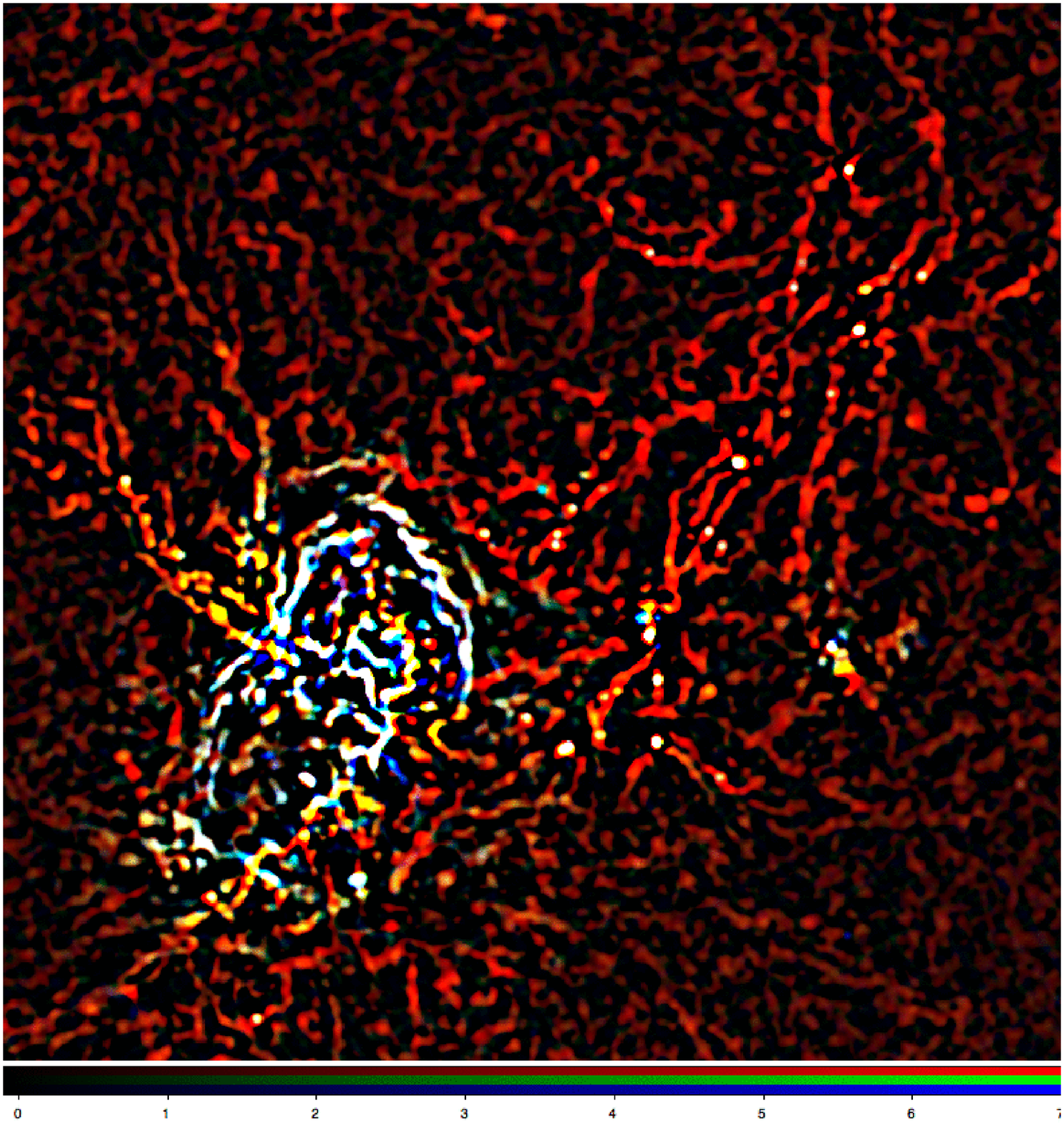}}                                
            \resizebox{0.5\hsize}{!}{\includegraphics{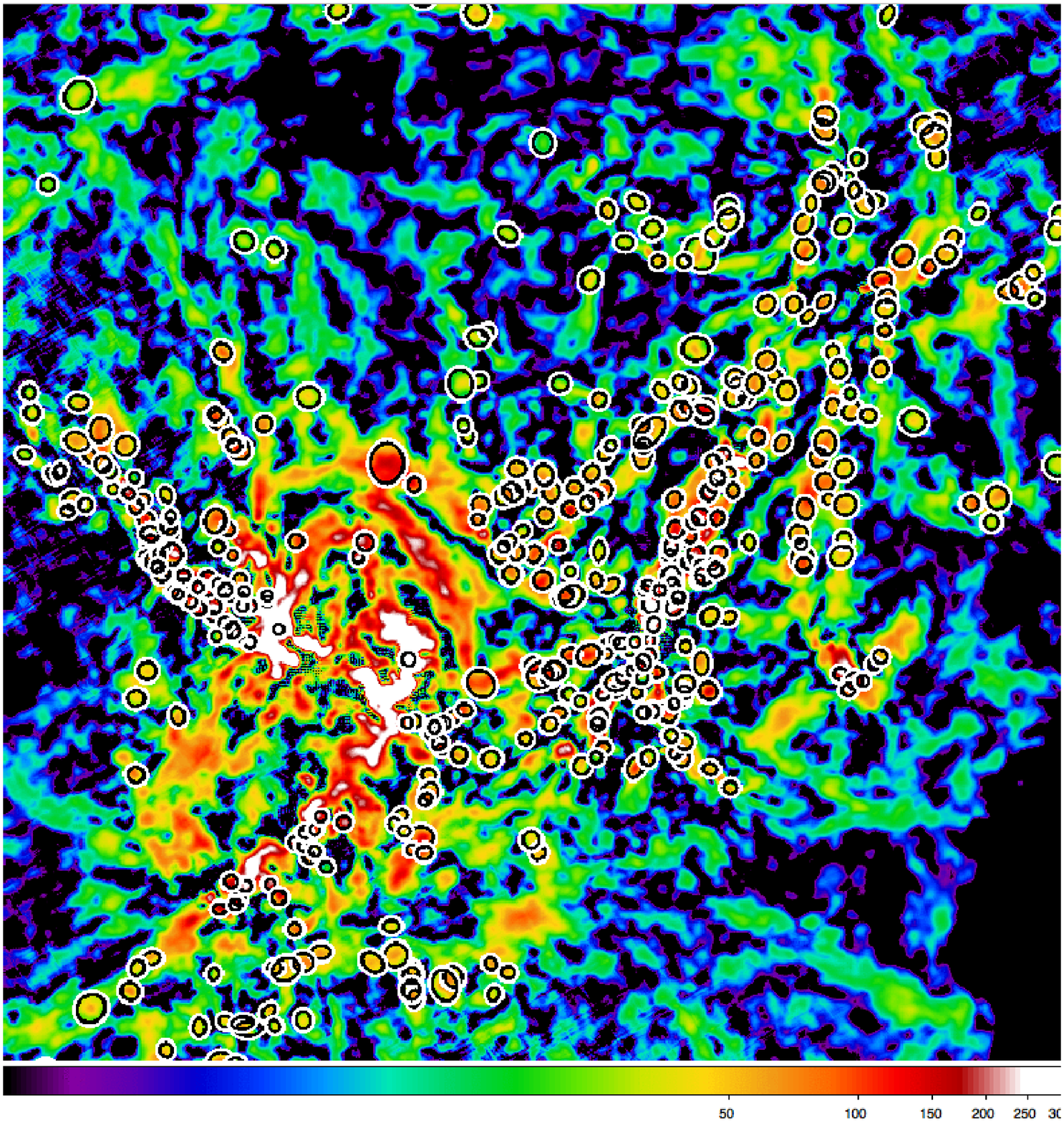}}}                                
\centerline{\resizebox{0.5\hsize}{!}{\includegraphics{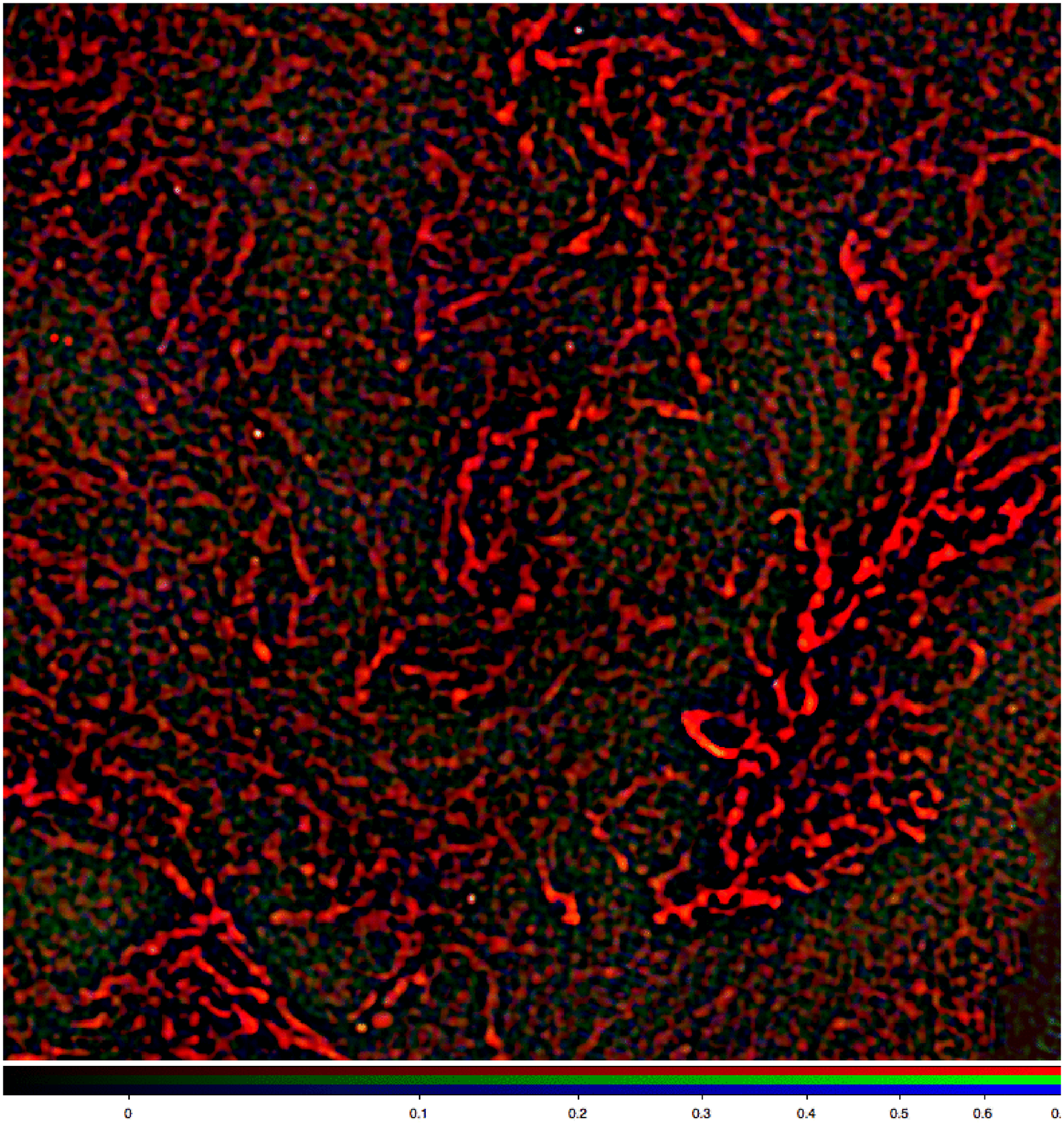}}
            \resizebox{0.5\hsize}{!}{\includegraphics{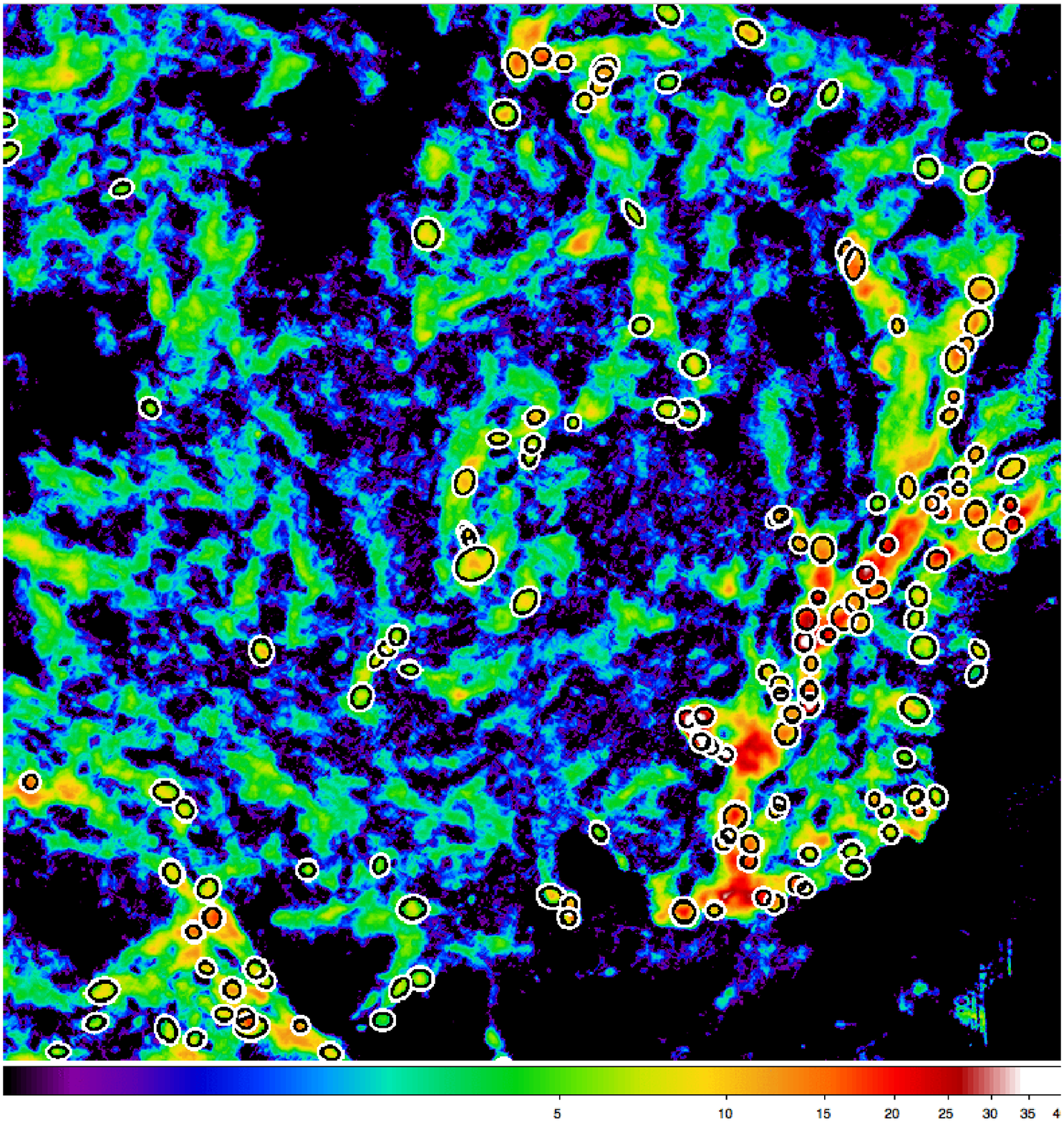}}}                                 
\caption{Composite 3-color images \emph{left} of the sub-fields of Aquila (\emph{top}) and Polaris (\emph{bottom}) produced from 
         the high-contrast ``single-scale'' decompositions (red comes from all SPIRE bands, green and blue correspond to the PACS 
         bands at 160 and 70\,{\um}, respectively); the sub-fields and decompositions are the same as in Fig.~\ref{figure1}. 
         Curvelet components (\emph{right}) extracted from the original SPIRE 350\,{\um} images overlaid with ellipses for selected 
         cores extracted by \emph{getsources} \citep[shown are starless cores: those detected at S/N\,$\ge$\,7.5 levels in
         at least two SPIRE bands and not detected in emission or detected in absorption in the PACS 70\,{\um} band, with SED dust 
         temperatures $T_{\rm d}\le 18$\,K; see also][]{Ko"nyves_etal2010,Andre_etal2010}. Most objects lie in the narrow and long
         filaments spanning orders of magnitude in intensity (MJy/sr).}
\label{figure2}
\end{figure*}
}


\section{Introduction}
\label{Introduction}

\emph{Herschel} \citep{Pilbratt_etal2010} provides the best opportunity to study the earliest stages of star formation and the
origin of the stellar initial mass function (IMF). Prestellar cores and young (Class 0) protostars emit the bulk of their
luminosities at wavelengths ~80--400\,{\um}, which makes the \emph{Herschel} imaging instruments PACS \citep{Poglitsch_etal2010}
and SPIRE \citep{Griffin_etal2010} perfect for performing a census of these objects down to ~0.01--0.1\,{\Msun} in the nearby
(distances $D \la$ 500 pc) molecular cloud complexes. Our \emph{Herschel} key project \citep[Gould Belt survey,][]{Andre_etal2010}
aims at probing the link between diffuse cirrus-like structures and compact cores with the main scientific goal to elucidate the
physical mechanisms of the formation of prestellar cores out of the diffuse medium, which is crucial for understanding the origin
of stellar masses.

During the science demonstration phase of \emph{Herschel} observations we imaged two large fields, the Aquila Rift and part of the
Polaris Flare (hereafter called Aquila and Polaris fields for brevity), with the aim of better understanding star formation in
extremely different environments. The high-quality far-infrared images are extremely rich in new information on the earliest phases
of star formation, which will be explored in full detail in the future. In this short paper we focus on merely one discovery made by
\emph{Herschel}: the impressive, ubiquitous network of filamentary structures in the interstellar medium (ISM) and its remarkably
close physical relationship with the objects that appear to form within the filaments.


\section{Observations and data reduction} 
\label{observations.and.data.reduction}

The PACS instrument is described by \citet{Poglitsch_etal2010}; the SPIRE instrument, its in-orbit performance, and its scientific
capabilities are described by \citet{Griffin_etal2010}, and the SPIRE astronomical calibration methods and accuracy are outlined by
\citet{Swinyard_etal2010}. Our \emph{Herschel} observations of the Aquila field are presented by \citet{Ko"nyves_etal2010}; we
describe here our observations of the Polaris field.

The entire Polaris field (8 square degrees) was observed on 2009 October 23, at 70, 160, 250, 350, and 500\,{\um} in the PACS/SPIRE
parallel mode at 60\,{\arcsec}s$^{-1}$, performing cross-linked scans in two orthogonal directions. The PACS raw data were reduced
with the HIPE 3.0.455 software with standard steps of the default pipeline. The baseline was subtracted by high-pass filtering with
a $\sim$1{\degr} median filter width (the full leg length was 2{\fdg}5). Multi-resolution median transform deglitching and second
order deglitching were applied, and final maps were computed by the HIPE's MADmap method. The SPIRE parallel-mode data were reduced
with HIPE 2.0 using pipeline scripts delivered with that version, modified to include data that were taken during the telescope's
turnaround at the scan ends. A median baseline was applied to the maps on individual scan legs and a ``naive mapper'' was used to
create images. The images were projected to the nominal 3{\farcs}2, 3{\farcs}2, and 6{\farcs}4 pixels for PACS and 6{\arcsec},
10{\arcsec}, and 14{\arcsec} pixels for SPIRE wavelengths. In our analysis, we adopted circular Gaussian point-spread functions
(PSFs) of 8{\farcs}4, 9{\farcs}4, 13{\farcs}5, 18{\farcs}1, 25{\farcs}2, and 36{\farcs}9 at the wavelengths (although the beams are
actually somewhat elongated). Before extracting objects, all images were resampled to the same 6{\arcsec} pixel size using
\emph{SWarp} \citep{Bertin_etal2002}.


\section{Source extraction method}
\label{source.extraction.method}

Here, we summarize the multi-scale, multi-wavelength source extraction algorithm that has been developed and extensively tested by
A.~M. at CEA Saclay. The main motivation for the development was the absence of any \emph{multi-wavelength} extraction technique
for use in our \emph{Herschel} projects. Full description of the method and code (called \textit{getsources}) as well as of the
benchmarking results for this and several other extraction techniques\footnote{The Gould Belt consortium is testing several source
extraction algorithms, such as \textit{gaussclumps} \citep{StutzkiGuesten1990}, \textit{clumpfind} \citep{WilliamsdeGeusBlitz1994},
\textit{sextractor} \citep{BertinArnouts1996}, \textit{derivatives} \citep{Molinari_etal2010}, and a few others (see Appendix A 
for a very brief summary), using simulated skies of various degrees of complexity.} will be given elsewhere 
\citep{Men'shchikov_etal2010}.

The main idea of the method used in this work is to analyze decompositions of original images (at each wavelength) across a wide
range of spatial scales separated by only a small amount (typically $\sim$ 5{\%}). Each of those ``\emph{single scales}'' are
cleaned of noise and background by iterating to appropriate cut-off levels, then re-normalized and summed up over all wavelengths
in a \emph{combined} single-scale detection image. The main advantage of this algorithm is in its multi-wavelength design: the
same combined detection image across all wavelengths eliminates the need of matching multiple catalogs obtained with different
angular resolutions and reduces associated measurement errors. Besides, fine spatial decomposition improves the detection of even
the faintest objects and aids in better deblending of crowded regions.

The decomposition is done by convolving the original images with circular Gaussians and subtracting them from one another (this can
be called \emph{successive unsharp masking}): 
\begin{equation}
I_{j}(\lambda) = G_{j-1} * I(\lambda) - G_{j} * I(\lambda)$  \,\,($j = 1, 2, ..., N),
\label{successive.unsharp.masking}
\end{equation}
where $I(\lambda)$ is the original image at a wavelength $\lambda$, $I_{j}(\lambda)$ are ``single-scale'' decompositions, $G_{j}$ 
are the smoothing Gaussians ($G_0$ is a two-dimensional delta function) with ${\rm PSF} \la {\rm FWHM} \la$ image size, 
${\rm FWHM}_{j} = f_{\rm s}\times {\rm FWHM}_{j-1}$, $f_{\rm s} > 1$ is a scale factor (usually $f_{\rm s}\approx 1.05$), PSF is 
the smallest beam over all $\lambda$, and the number of scales $N$ depends on the value of $f_{\rm s}$.

To separate the signals of objects from the noise and background contributions, we iterate to $5\sigma_{j}(\lambda)$ cut-off levels
in each of $I_{j}(\lambda)$ (the standard deviation $\sigma_{j}(\lambda)$ is computed outside the objects) and zero out all pixels
with intensities below that level, producing clean images $I_{j}^{\rm c}(\lambda)$. An advantage of this \emph{single-scale
cleaning} is that the noise contribution depends very significantly on the scale (e.g., at large scales the small-scale noise is
diluted, whereas large objects are best visible). In effect, in a full clean image $I^{\rm c}(\lambda)$\,=$\sum_{j} I_{j}^{\rm
c}(\lambda)$ one would see structures deeper than in $I(\lambda)$. The above algorithm is applied to images at each band separately
and the clean images are used to produce \emph{combined} single-scale \emph{detection} images $I_{j}^{\rm c}$\,=$\sum_{\lambda}
f\times I_{j}^{\rm c}(\lambda)$ after normalization to a similar maximum intensity\footnote{At this detection step, it is neither
possible nor necessary to properly preserve the spectral behavior of sources at single scales in the combined detection image, as
the processing is done at the entire image level before finding objects; spectral properties of the latter will be determined from
the original images at the measurement step.}. Objects in the combined detection image are identified by tracking their appearance
and ``evolution'' from small to large scales, and their \emph{footprints} are determined. The objects are background-subtracted and
deblended, and their sizes and fluxes are measured in the original observed images at each wavelength. For more details, we refer
interested readers to Appendix A.


\section{Results and discussion}
\label{results.and.discussion}

To derive properties of the compact objects in the Aquila and Polaris fields, we performed source extraction with
\textit{getsources} (Sect.~\ref{source.extraction.method}) and made a careful analysis of the resulting extraction catalogs,
selecting most reliable objects and determining core mass function for both fields \citep{Ko"nyves_etal2010,Andre_etal2010}. To
assess completeness levels of our observations, we created the synthetic skies with the actual observed background of our images
(after removing all extracted objects) and populated them with cores and protostars from spherical radiative transfer models
\citep{Men'shchikov_etal2010}, then performed extractions on the skies.

The filamentary structures are clearly visible in the left panels of Fig.~\ref{figure1}, where both the Aquila and Polaris fields
are shown as 3-color composites created from the images at 500, 160, and 70\,{\um}. With red color, the images highlight strong
density enhancements and low temperatures, whereas the material within two bright hot regions (HII region W\,40 in the north and
around MWC\,297 and Sh\,2-62 stellar group in the south) of the Aquila field is visible in green, blue, or white. From this one
could immediately conclude where to look for prestellar cores, and it is indeed the red filaments where the reddest cores are
visible. The colors are not very pronounced in the Polaris field, suggesting much smaller temperature and density variations across
the field, even in its densest filaments. To better visualize and characterize the filaments, we utilized several approaches.

High-contrast images of the cores and filaments in both fields were obtained as part of our source extraction technique by applying
the successive unsharp masking (Eq.~\ref{successive.unsharp.masking}) with $f_{\rm s}$=1.05, $G_1$=9{\farcs}39, and $N$=99. In
particular, the right panels in Fig.~\ref{figure1} show that the visibility of the very narrow and long filaments is dramatically
improved in some of our ``single-scale'' images (which effectively select a narrow range of scales, with both larger and smaller
scales filtered out). For the sake of illustration, a 40{\arcsec} scale (in which the structures are best visible) is displayed in
Fig.~\ref{figure1}. The compact objects are populating the filaments almost like pearls on threads in a necklace, giving them a
nodular, lumpy appearance. The same decomposition at all \emph{Herschel} wavelengths proved that the same structures are seen
across the entire range, from 70 to 500\,{\um}. There are, however, noticeable changes in the appearance of the filaments in Aquila
(around W\,40) at the shortest wavelengths (Fig.~\ref{figure2}, upper left), presumably due to increasing opacity and to changes in
temperature of some of the filaments (complex three-dimensional structures).

We also extracted filamentary structure from the observed images with the MCA software \citep[morphological component
analysis,][]{Starck_etal2004}. The idea of the method is to morphologically decompose a signal into its building blocks, which can
be represented by isotropic wavelets, ridgelets, or curvelets. Filamentary structures were separated from the more isotropic
features by applying both the curvelet and wavelet transforms in 100 iterations (with exponentially decreasing thresholds) to
achieve convergence in both the wavelet and curvelet representations of the original images. The curvelet component
\citep[Fig.~\ref{figure2}, right panels; see also][]{Andre_etal2010} shows filaments very similar to those displayed in
Fig.~\ref{figure1} and left panels of Fig.~\ref{figure2}; we also get analogous results when processing the column density images
\citep[the latter are shown in][]{Ko"nyves_etal2010}.



Measurements of the thickness of several high-contrast filaments in our column density maps (obtained with a uniform 39{\arcsec}
resolution of the 500\,{\um} band) give 48{\arcsec} for Aquila and 66{\arcsec} for Polaris (FWHM), uncertain to within 12{\arcsec};
when deconvolved, they are 28{\arcsec} and 53{\arcsec}. If we adopt the distances to the regions of 260\,pc
\citep{Straizys_etal2003,Bontemps_etal2010} and 150\,pc \citep{HeithausenThaddeus1990}, the deconvolved linear widths are $\sim$
7300\,AU and 8000\,AU. Measurements of the widths at several locations across a few well-behaved filaments (those not blended with
the other filaments) in the observed images resulted in the values of 30, 36, 42, 42, 54{\arcsec} (for Aquila at the PACS+SPIRE
wavelengths) and 63, 64, 69{\arcsec} (for Polaris at the SPIRE wavelengths). The filaments are more resolved at PACS wavelengths
and the values would imply average deconvolved widths of 35{\arcsec} for Aquila and 59{\arcsec} for Polaris (9000\,AU in both
fields). The largest FWHM sizes of the extracted cores are comparable to the filaments' widths (within $\sim$50{\%}) for both
fields, confirming the visual impression that the objects are largely confined to the filaments they are embedded in. There seems
to be a wide distribution of the lengths of the filaments; some of them are as long as $\sim$0{\fdg}5 (Figs.~\ref{figure1},
\ref{figure2}), which would correspond to a few pc. It is clear, however, that one needs to perform a more systematic and complete
study of the filaments' properties; that will be a subject of a future work.

The Aquila and Polaris fields present very different physical conditions; the former shows very active star formation, while the
latter displays almost none. Indeed, we find many extremely cold and dense filaments ($T_{\rm{d}}\approx$ 7.5--15\,K,
$N_{\rm{H_2}}\approx 5\times 10^{20}$--$1.4\times 10^{23}$\,cm$^{-2}$) in Aquila; most of them are gravitationally unstable and
fragmented into several hundreds of prestellar cores \citep{Andre_etal2010}. In contrast, the Polaris field contains warmer
filaments with much lower column densities ($T_{\rm{d}}\approx$ 10--15\,K, $N_{\rm{H_2}}\approx 3\times 10^{20}$--$8.6\times
10^{21}$\,cm$^{-2}$) and we find no clear examples of prestellar cores \citep{Andre_etal2010,Ward-Thompson_etal2010}. To illustrate
the very close relationship between the filaments and detected objects, we overplot locations of the latter on the curvelet
transform for both fields. An inspection of Fig.~\ref{figure2} shows that all objects detected in Aquila and Polaris are indeed
located within very narrow filaments. Our maps of a stability parameter of the filaments (their mass per unit length) suggest that
the Polaris filaments are stable, while many dense filaments in Aquila are gravitationally unstable \citep{Andre_etal2010}.

Several possible models for the formation of filamentary cloud structures have been proposed in the literature. In particular,
numerical simulations of supersonic magnetohydrodynamic turbulence in weakly magnetized clouds always generate complex systems of
shocks, which fragment the gas into high-density sheets, filaments, and cores \cite[e.g.][]{Padoan_etal2001}. Filamentary
structures are also produced in turbulent simulations of more strongly magnetized molecular clouds, including the effects of
ambipolar diffusion \citep{NakamuraLi2008}. Another possibility is that the filaments originate from gravitational fragmentation of
sheet-like cloud layers \citep{Nagai_etal1998}. Interestingly, filaments can result from purely gravitational amplification of a
random field of initial density fluctuations as is the case in large cosmological simulations \citep[e.g.][]{Ocvirk_etal2008}. Our
measurements of the radial profiles of selected filaments in Aquila and Polaris suggest that magnetic fields may play an important
role in shaping the structure of the observed filaments, if not in forming the filaments. Indeed, they indicate that the density
falls off approximately as $\rho\propto r^{-1.5}$ (in Aquila) or $\rho\propto r^{-2}$ (in Polaris) away from the axes of the
filaments. In both cases, the measured radial density profiles are much shallower than the steep $\rho\propto r^{-4}$ profile of
unmagnetized hydrostatic filaments \citep{Ostriker1964}, as already pointed out by \citet{Lada_etal1999} and
\citet{JohnstoneBally1999} for two specific filaments. Models of unmagnetized equilibrium filaments are therefore inconsistent with
observations. By contrast, models of equilibrium filaments with primarily toroidal or helical magnetic fields can account for
radial profiles ranging from $\rho\propto r^{-1}$ to $\rho\propto r^{-2}$, in agreement with observations \citep{FiegePudritz2000}.

Filaments are seen in numerous astronomical images and the filamentary structure of molecular clouds has been known for some time
\citep[e.g.][]{SchneiderElmegreen_1979,Goldsmith_etal2008}. However, \emph{Herschel} with its combination of the far-IR resolution
and sensitivity \emph{directly} reveals the filaments in which the cores are forming. Our first observations demonstrate the
fascinating, omnipresent filamentary density structure of the cold ISM and its extraordinarily intimate physical relationship with
the objects that form within the filaments. They enable us to suggest that in general dense cores may originate in a process of
fragmentation of long, thin filaments, presumably formed as a result of an interplay between gravity, interstellar turbulence, and
magnetic fields \citep[see also discussion in][]{Andre_etal2010}. To unravel the role and relative importance of these processes,
one needs additional kinematic and polarization information \citep[cf.][]{Matthews_etal2001}; these follow-up observations are
planned.


\begin{acknowledgements}

PACS has been developed by a consortium of institutes led by MPE (Germany) and including UVIE (Austria); KU Leuven, CSL, IMEC
(Belgium); CEA, LAM (France); MPIA (Germany); INAF-IFSI/ OAA/OAP/OAT, LENS, SISSA (Italy); IAC (Spain). This development has been
supported by the funding agencies BMVIT (Austria), ESA-PRODEX (Belgium), CEA/CNES (France), DLR (Germany), ASI/INAF (Italy), and
CICYT/MCYT (Spain). SPIRE has been developed by a consortium of institutes led by Cardiff Univ. (UK) and including Univ. Lethbridge
(Canada); NAOC (China); CEA, LAM (France); IFSI, Univ. Padua (Italy); IAC (Spain); Stockholm Observatory (Sweden); Imperial College
London, RAL, UCL-MSSL, UKATC, Univ. Sussex (UK); Caltech, JPL, NHSC, Univ. Colorado (USA). This development has been supported by
national funding agencies: CSA (Canada); NAOC (China); CEA, CNES, CNRS (France); ASI (Italy); MCINN (Spain); SNSB (Sweden); STFC
(UK); and NASA (USA).

\end{acknowledgements}


\bibliographystyle{aa}
\bibliography{aamnem99,filaments}


\Online
\begin{appendix}

\section{Extraction techniques}
\label{appendix}

\subsection{Existing source extraction algorithms}
\label{extraction.algorithms}

Here we summarize (very briefly) the concepts of different techniques, to place \emph{getsources} described in
Sect.~\ref{source.extraction.method} in a wider context. The algorithms trying to solve the same problem of source extraction
originated from different ideas. Note that they have also been developed (oriented) for use in different areas of astronomy, thus 
their performance for a specific project must be carefully tested before an appropriate method can be chosen.

\cite{StutzkiGuesten1990}'s \emph{gaussclumps} (originally created for position-velocity cubes) fits a Gaussian profile to the
brightest peak, subtracting the fit from the image, then fitting a new profile to the brightest peak in the image of residuals,
iterating until some termination criteria are met. \cite{WilliamsdeGeusBlitz1994}'s \emph{clumpfind} contours an image at a number
of levels, starting from the brightest peak in the image and descending down to a minimum contour level, marking as clumps along
the way all connected areas of pixels that are above the contour level. \cite{BertinArnouts1996}'s \emph{sextractor} estimates and
subtracts background, then uses thresholding to find objects, deblends them if they overlap, and measures their positions and sizes
using intensity moments. CUPID\footnote{CUPID is a source extraction software package developed by the STARLINK team for use with
the SCUBA2 surveys; it is a general wrapper to which additional methods can be added. See documentation:
http://docs.jach.hawaii.edu/star/sun255.htx/sun255.html}'s \emph{reinhold} identifies pixels within the image which mark the edges
of clumps of emission, producing a set of rings around the clumps. After cleaning noise effects on the edges, all pixels within
each ring are assumed to belong to a single clump. CUPID's \emph{fellwalker} ascends image peaks by following the line of the
steepest ascent, considering every pixel in the image as a starting point for a walk to a significant peak, marking along the way
all visited pixels with a clump identifier. \cite{Motte_etal2007}'s \emph{mre-gcl} combines cloud filtering techniques based on
wavelet multi-resolution algorithms \citep[e.g.,][]{StarckMurtagh2006} with \emph{gaussclumps}. \cite{Molinari_etal2010}'s
\emph{derivatives} analyzes multi-directional second derivatives of the original image and performs curvature thresholding to 
isolate compact objects, then fits variable-size elliptical Gaussians (adding also a planar background) at their positions. 
Another method that defines cores in terms of connected pixels is \emph{csar}, which was developed for use with BLAST and 
\emph{Herschel} (Harry et al. 2010, in preparation).

Whereas \emph{clumpfind}, \emph{reinhold}, \emph{fellwalker}, and \emph{csar} merely partition the image between objects not
allowing them to overlap, \emph{gaussclumps}, \emph{sextractor}, and \emph{mre-gcl} can deblend overlapping objects, which is quite
essential for obtaining correct results in crowded regions. None of the methods was designed to handle multi-wavelength data,
making it necessary to match the catalogs obtained at different wavelegths using an association radius as a parameter.

\subsection{More details on the new method}
\label{getsources.details}

In \emph{getsources} the extraction of objects is performed in each of the combined detection images by going from the smallest to 
the largest scales and finding segmentation masks of the objects at each scale using the \emph{tint fill} algorithm 
\citep{Smith_1979}\footnote{Available at http://portal.acm.org/citation.cfm?id=800249.807456}. 
The masks are the areas of connected pixels in a segmentation image, and the algorithm fills the pixels' values with the number of a
detected object and allows tracking of all pixels belonging to the object across all scales. The segmentation masks expand toward
larger scales, and the evolution of each object's mask is followed, as is appearance of new objects at any scale and disappearance
of those which become too faint at the current and larger scales. When two or more objects touch each other in a single-scale
image, the segmentation masks are not allowed to overlap, but overlapping does happen between objects of different scales. The
largest extent of any source defines its \emph{footprint}, and this is determined at the scale where the object's contrast
above the cut-off level is at maximum. The scale itself provides an initial estimate for the object's FWHM size.

The positions of sources are computed from the first moments of intensities in a combined detection image at a range of single
scales, from where an object first appeared and to the scale twice as large. The objects' sizes are computed from the first and
second intensity moments in the original \emph{background-subtracted} image. The background subtraction is done by linearly
interpolating pixel intensities off the observed image under the footprints, in the four main directions (two axes and two
diagonals), based on the pixel values just outside the footprints. Our iterative deblending algorithm employs two-dimensional
shapes with peak intensities and sizes of the extracted objects in order to divide the intensity of a pixel between surrounding
objects according to the fraction of the shapes' intensities at the pixel. For the shapes we adopted a two-dimensional analog of
the Gaussian-like function $f_0(1+(r/r_0)^2)^{-\alpha}$ \citep{Moffat_1969} with $\alpha=10$.

The end result of the processing is an extraction catalog (one line per object) containing coordinates of all detections
(independent of $\lambda$) and estimates of the objects' S/N ratios, peak and total fluxes (with their uncertainties), and sizes
and orientations for each wavelength. In addition, \emph{getsources} produces catalogs of all possible colors, as well as the
azimuthally-averaged intensity profiles (their full, background-subtracted, and deblended versions) and deblended images for each
object.

\end{appendix}

\end{document}